**Single-Particle Tunneling in Doped Graphene-Insulator-Graphene Junctions**

R. M. Feenstra[*]
Dept. Physics, Carnegie Mellon University, Pittsburgh, PA 15213

Debdeep Jena[†]
Dept. Electrical Engineering, University of Notre Dame, Notre Dame, IN 46556

Gong Gu[‡]
Dept. Electrical Engineering and Computer Science, University of Tennessee, Knoxville, TN 37996

**Abstract**

The characteristics of tunnel junctions formed between n- and p-doped graphene are investigated theoretically. The single-particle tunnel current that flows between the two-dimensional electronic states of the graphene (2D-2D tunneling) is evaluated. At a voltage bias such that the Dirac points of the two electrodes are aligned, a large resonant current peak is produced. The magnitude and width of this peak are computed, and its use for devices is discussed. The influences of both rotational alignment of the graphene electrodes and structural perfection of the graphene are also discussed.

**I. Introduction**

Two-dimensional (2D) electron systems have played a very important role in the development of electronic devices, including metal-oxide-semiconductor field-effect transistors (MOSFETs) made from silicon and high electron mobility transistors (HEMTs) made from III-V semiconductor heterostructures.[1] One lesser-known device utilizing 2D electron gases (2DEGs) is a tunnel junction between two such gases, i.e. 2D-2D tunneling. Prior investigations of 2D-2D tunneling have been carried out on coupled electron gas systems in closely placed quantum wells in AlGaAs/GaAs heterostructures.[2,3] Considering the case of unequal doping between the 2DEGs, it was demonstrated experimentally that, at a voltage bias corresponding to aligned band structures of the 2D systems, a large, sharp peak in the tunnel current occurs. We refer to this peak as a *resonant peak* in the tunneling. It was argued in the prior work that the width of this peak was temperature independent[2,3] (except possibly from inelastic effects).

With the advent of a new 2D electronic system, graphene, it is worthwhile to consider how 2D-2D tunneling could be employed in this system. In this work we theoretically investigate that question, examining graphene-insulator-graphene (GIG) tunnel junctions.

---

[*] feenstra@cmu.edu
[†] djena@nd.edu
[‡] ggu1@utk.edu

We focus in particular on the situation when the graphene sheets have unequal doping, e.g. one is n-type (electron doped) and the other is p-type (hole doped). We derive formulas for the voltage-dependence of the current, results that were not obtained, to our knowledge, in any prior 2D-2D tunneling work (although Ref. [4] provided a step in this direction). A large current peak occurs at the voltage when the band structures of the graphene sheets are energetically aligned (and also the graphene sheets are rotationally aligned in real space), and this peak is characterized in terms of its magnitude and width. We consider finite-size areas for the graphene sheets, as might occur physically due to the limited size of structurally perfect regions in the graphene, something that we denote by a "structural coherence length" $L$. We find that the magnitude of the resonant current peak is proportional to the electrode area times $L$ and its width is proportional to $1/L$. Compared to other nonresonant aspects of the current, very high degrees of nonlinearity in the current-voltage (*I-V*) relationship remain even for values of $L$ as low as 100 nm or less.

Clearly this sort of highly-nonlinear *I-V* relationship has potential applications for electronic devices. The sharp resonant current peak at small voltages presents a compelling case for being integrated into a three-terminal device where the third terminal (a capacitive gate) can move the effective bias on and off the resonance condition, thereby enabling logic operations. In fact, precisely this sort of device, a *BiSFET* utilizing a graphene bilayer, has been recently proposed as a low-power building block for logic operations.[5,6] The operation of that device however is based on many-body excitonic condensate effects, which will be observed only below a certain characteristic critical temperature.[5,6] Our work is for single-particle tunneling, where the condensate is not required and hence there is no critical temperature. On the other hand, devices utilizing a single-particle tunneling resonance *do* require, at least for optimal performance, rotational alignment of the graphene electrodes and a well-ordered insulating layer (to minimize momentum scattering), things that are not needed for the excitonic mechanism of the BiSFET. Both types of devices are quite impervious to effect of thermal broadening, and both devices are also intrinsically fast since they rely on tunneling.

In Section II we present our general theoretical method using the Bardeen transfer Hamiltonian approach, followed by applications of that to both undoped and doped GIG junctions. The contributions to the current are described analytically, with finite-size effects being considered in particular. Numerical results for the current are provided in Section III, and in Section IV we discuss the results and briefly consider possible fabrication of GIG junctions and extension to three-terminal devices. The paper is summarized in Section V.

**II. Theory**
**A. Qualitative considerations**

The nonlinear *I-V* characteristic of a GIG junction with complementary doping in the graphene electrodes is easily seen by considering the states available for tunneling, as illustrated in Fig. 1. We assume that the left-hand electrode is n-doped and the right-hand electrode is p-doped, with chemical potentials (Fermi levels) $\mu_L = E_{DL} + \Delta E_L$ and $\mu_R = E_{DR} - \Delta E_R$ for specific $\Delta E_L$ and $\Delta E_R$, where $E_{DL}$ and $E_{DR}$ are the respective Dirac points. For simplicity we assume $\Delta E_L = \Delta E_R \equiv \Delta E > 0$. For applied voltage bias $V$ between the electrodes we have $\mu_L - \mu_R = eV$. It is important to note that, for our situation of graphene electrodes, the value of $\Delta E$ will depend not only on the doping of the electrodes but also on the applied bias $V$ and the geometric capacitance $C$ of the GIG junction (due to the quantum capacitances of the graphene electrodes).[7] This dependence of $\Delta E$ is described in Section II(E) below, and for the present discussion we take $\Delta E$ to be a fixed quantity.

Let us first consider the nonresonant case when the band structures are not aligned, $eV \neq 2\Delta E$, as in Figs. 1(a) and 1(b) for voltage ranges of $eV < 2\Delta E$ and $eV > 2\Delta E$, respectively. Then, given the requirement of momentum conservation (for large area, rotationally aligned graphene electrodes, and neglecting scattering in the insulator), there is only a single ring of **k**-points that can satisfy that, located at an energy midway between the Dirac points as shown in Figs. 1(a) and 1(b). The circumference of these rings varies linearly with voltage, producing a linear dependence of the current on voltage as pictured in the *I-V* curve of Fig. 1(d).

Now we turn to the resonant situation, with $eV = 2\Delta E$. As pictured in Fig. 1(c), there are states existing over all energies that satisfy the requirement of **k**-conservation. The resulting current is relatively large, scaling superlinearly with the area of the electrodes (since the number of states involved increases with the area). This current is pictured as the upwards pointing arrow in Fig. 1(d). As will be shown in the following Sections, this resonant peak in the current has an amplitude that scales as the area of the electrode times a "structural coherence length" $L$, with $L$ being just $\sqrt{A}$ for a perfectly crystalline graphene sheet with area $A$, or a typical length between defects in the sheet, whichever is less. The width of the resonance peak scales as $1/L$.

**B. Formalism**

We compute tunnel currents using the Bardeen transfer Hamiltonian approach,[8,9]

$$I = g_S g_V e \sum_{\alpha,\beta} \left\{ \frac{1}{\tau_{\alpha\beta}} f_L(E_\alpha)[1 - f_R(E_\beta)] - \frac{1}{\tau_{\beta\alpha}} f_R(E_\beta)[1 - f_L(E_\alpha)] \right\}, \quad (1)$$

where $\alpha$ and $\beta$ label states in the left- (*L*) and right-hand (*R*) electrodes with energies of $E_\alpha$ and $E_\beta$ respectively, $g_S = 2$ is the spin degeneracy and $g_V$ is the valley degeneracy, $\tau_{\alpha\beta}^{-1}$ and $\tau_{\beta\alpha}^{-1}$ are the tunneling rates for electrons going $L \rightarrow R$ or $R \rightarrow L$ respectively, and $f_L$ and $f_R$ are Fermi occupation factor for the left and right-hand electrodes, $f_L(E) = \{1 + \exp[(E - \mu_L)/k_B T]\}^{-1}$ and $f_R(E) = \{1 + \exp[(E - \mu_R)/k_B T]\}^{-1}$. The tunneling rates are given by

$$\frac{1}{\tau_{\alpha\beta}} = \frac{2\pi}{\hbar}|M_{\alpha\beta}|^2 \delta(E_\alpha - E_\beta) = \frac{1}{\tau_{\beta\alpha}}, \quad (2)$$

where

$$M_{\alpha\beta} = \frac{\hbar^2}{2m}\int dS \left( \Psi_\alpha^* \frac{d\Psi_\beta}{dz} - \Psi_\beta \frac{d\Psi_\alpha^*}{dz} \right) \quad (3)$$

is the matrix element for the transition with $m$ being the free electron mass and $\Psi_\alpha(\mathbf{r},z)$ and $\Psi_\beta(\mathbf{r},z)$ being the wavefunctions of the left- and right-hand electrodes, respectively. The surface integral in Eq. (3) is evaluated over a plane located midway between the two electrodes. The current thus becomes

$$I = g_V \frac{4\pi e}{\hbar} \sum_{\alpha,\beta} |M_{\alpha\beta}|^2 [f_L(E_\alpha) - f_R(E_\beta)] \delta(E_\alpha - E_\beta). \quad (4)$$

We consider the situation for graphene, with two identical atoms, labeled 1 and 2, per unit cell. The wavefunction for wavevector $\mathbf{k}$ can be written in terms of basis functions $\Phi_{j\mathbf{k}}$ ($j=1,2$) on each atom as $\Psi(\mathbf{r},z) = \chi_1(\mathbf{k})\Phi_{1\mathbf{k}}(\mathbf{r},z) + \chi_2(\mathbf{k})\Phi_{2\mathbf{k}}(\mathbf{r},z)$. The basis functions themselves have Bloch form, $\Phi_{j\mathbf{k}}(\mathbf{r},z) = \exp(i\mathbf{k}\bullet\mathbf{r})u_{j\mathbf{k}}(\mathbf{r},z)/\sqrt{A}$ where $u_{j\mathbf{k}}(\mathbf{r},z)$ is a periodic function and $A$ is the area of the electrode. These periodic functions are of course localized around the basis atoms (i.e. as $2p_z$ orbitals) of each graphene electrode, but in the plane midway between the electrodes the functions are spread out. Thus, as a function of the 2D radial coordinate $\mathbf{r}$ in this plane, the $u_{j\mathbf{k}}(\mathbf{r},z)$ functions will vary only weakly and that dependence will not largely affect the integral. (Importantly, nodes in the wavefunction are included in the $\chi_1(\mathbf{k})$ and $\chi_2(\mathbf{k})$ factors, specified below).

We therefore approximate the tunneling matrix element, incorporating the small influence of the radial dependence of the $u_{j\mathbf{k}}(\mathbf{r},z)$ into numerical constants, and assuming for the $z$-dependence the usual tunneling form $2\kappa e^{-\kappa d}/D$ where $d$ is the separation of the electrodes, $\kappa$ is the decay constant of the wavefunctions in the barrier,[10] and $D$ is a normalization constant for the $z$-part of the wavefunctions in the graphene, i.e., approximately equal to an interplanar separation in graphite.[9,11] (For very thin barriers this form for the $z$-dependence may not be so appropriate, but its order of magnitude should still be correct). For example, for a term in Eq. (3) involving the $u_{1\mathbf{k}_L}(\mathbf{r},z)$ part of $\Psi_\alpha(\mathbf{r},z)$ and the $u_{1\mathbf{k}_R}(\mathbf{r},z)$ part of $\Psi_\beta(\mathbf{r},z)$ we assume

$$\int dS\, e^{i(\mathbf{k}_R - \mathbf{k}_L)\bullet\mathbf{r}}\left( u_{1\mathbf{k}_L}^* \frac{du_{1\mathbf{k}_R}}{dz} - u_{1\mathbf{k}_R} \frac{du_{1\mathbf{k}_L}^*}{dz} \right) \approx \frac{2\kappa}{D}e^{-\kappa d}\, u_{11}^2 \int dS\, e^{i(\mathbf{k}_R - \mathbf{k}_L)\bullet\mathbf{r}} \quad (5)$$

where $u_{11}$ is a constant of order unity. This constant is also taken to have no dependence on $\mathbf{k}_L$ or $\mathbf{k}_R$, i.e., employing an effective-mass approximation in which the periodic functions are evaluated at the band extrema. In the same sense, we replace the total wavevector by $\mathbf{k}_0 + \mathbf{k}$ where $\mathbf{k}_0$ is the wavevector of the band extrema and $\mathbf{k}$ is the component of the wavevector relative to that. The term involving $u_{2\mathbf{k}_L}(\mathbf{r},z)$ and $u_{2\mathbf{k}_R}(\mathbf{r},z)$ is approximated in an identical way, yielding constant $u_{22}$ but with $u_{22} = u_{11}$ since the atoms in the unit cell are identical. Cross terms yields constants $u_{12} = u_{21}$ which also have order unity (though with magnitude likely to be less than $u_{11}$). For the $\chi_1(\mathbf{k})$ and $\chi_2(\mathbf{k})$ factors, they have the values well known for graphene in a nearest-neighbor tight-binding approximation[12]

$$\begin{bmatrix} \chi_1 \\ \chi_2 \end{bmatrix} = \frac{1}{\sqrt{2}} \begin{bmatrix} e^{\mp i\theta_\mathbf{k}/2} \\ se^{\pm i\theta_\mathbf{k}/2} \end{bmatrix} \tag{6}$$

where $\theta_\mathbf{k}$ is the angle of the relative wavevector, the upper sign is for a band extremum at the K point of the Brillouin zone and the lower for a K' point, and with $s = +1$ for the conduction band (CB) or $-1$ for the valence band (VB).

For rotationally misaligned graphene electrodes, we consider tunneling between bands in the respective electrodes with extrema that differ by a vector $\mathbf{Q}$, i.e. $\mathbf{k}_{0,R} = \mathbf{k}_{0,L} + \mathbf{Q}$ with a $\mathbf{Q}$ vector such that $|\mathbf{k}_{0,L}| = |\mathbf{k}_{0,R}| = |\mathbf{k}_{0,L} + \mathbf{Q}| = 4\pi/3a$ (the magnitude of the wavevector at the K and K' points) where $a = 0.2464$ nm is the graphene lattice constant. The matrix element is then found to be

$$M_{\alpha\beta} = \frac{\hbar^2 \kappa}{2AmD} e^{-\kappa d} g_\omega(\theta_L, \theta_R) \int dS \, e^{i\mathbf{Q}\cdot\mathbf{r}} e^{i(\mathbf{k}_R - \mathbf{k}_L)\cdot\mathbf{r}} \tag{7}$$

where

$$g_\omega(\theta_L, \theta_R) = u_{11}^2 \left( e^{i(\theta_L \mp \theta_R')/2} + s_L s_R e^{-i(\theta_L \mp \theta_R')/2} \right) \\ + u_{12}^2 \left( s_R e^{i(\theta_L \pm \theta_R')/2} + s_L e^{-i(\theta_L \pm \theta_R')/2} \right) \tag{8}$$

with the upper sign used for tunneling between like valleys (i.e. K to K, or K' to K') and the lower sign for unlike valleys (K to K', or K' to K), where $\theta_R' \equiv \theta_R + \omega$ with $\omega = 2\sin^{-1}(3aQ/8\pi)$ being the misalignment angle between the electrodes, and where we have defined $\theta_L \equiv \theta_{\mathbf{k}_L}$ and $\theta_R \equiv \theta_{\mathbf{k}_R}$. For the case of nonzero $\mathbf{Q}$ (nonzero $\omega$), the values of the $u_{ij}$ constants will change, but as argued above these constants have little effect on the resulting current (at least for moderately thick barriers) so we do not explicitly consider that change. We note that the $g_\omega$ factor of Eq. (8) has only a relatively small influence on the final results for the tunnel current, but it is nevertheless included in our analysis for completeness.

For rotationally aligned electrodes we have $\mathbf{Q} = 0$, so that the integral on the right-hand side of Eq. (7) approaches the delta-function $\delta(\mathbf{k}_R - \mathbf{k}_L)$ for $A \to \infty$. Of particular interest in our discussion below is the situation for finite-area tunnel junctions, in which case we will want to evaluate this integral for moderate-sized values of *A*. It is convenient to work in terms of the square of the integral from the right-hand side of Eq (7),

$$\Lambda(\Delta \mathbf{k}) \equiv \left| \frac{1}{A} \int dS \, e^{i \Delta \mathbf{k} \bullet \mathbf{r}} \right|^2 \tag{9}$$

with $\Delta \mathbf{k} = \mathbf{k}_R - \mathbf{k}_L$, and where for large $A$, $\Lambda(\Delta \mathbf{k}) \to \delta^2_{\mathbf{k}_L, \mathbf{k}_R} = \delta_{\mathbf{k}_L, \mathbf{k}_R}$. In Section II(D) we consider other formulas and/or approximations to $\Lambda(\Delta \mathbf{k})$ as appropriate to the case when *A* is not so large. Incorporating Eqs. (7) and (8) into (4), and with $g_V = 2$ for graphene, we arrive at the expression for the current (with states labeled by $\mathbf{k}_L$ or $\mathbf{k}_R$)

$$I = \frac{8\pi e}{\hbar} \left( \frac{\hbar^2 \kappa e^{-\kappa d}}{2mD} \right)^2 \sum_B \sum_{\mathbf{k}_L, \mathbf{k}_R} |g_0(\theta_L, \theta_R)|^2 \\ \times \left[ f_L(E_{\mathbf{k}_L}) - f_R(E_{\mathbf{k}_R}) \right] \delta(E_{\mathbf{k}_L} - E_{\mathbf{k}_R}) \Lambda(\Delta \mathbf{k}) \tag{10}$$

where $g_0(\theta_L, \theta_R)$ is defined by Eq. (8) with $\omega = 0$.

The sum over *B* in Eq. (10) indicates the different regimes of relative band alignments between the electrodes, labeled I, II, or III in Fig. 2, that must be considered in evaluating the current. For example, in region I we have $E_{\mathbf{k}_L} = E_{DL} + \hbar v_F k_L$ and $E_{\mathbf{k}_R} = E_{DR} + \hbar v_F k_R$ where $v_F$ is the Fermi velocity ($\approx c/300$), so that the argument of the energy δ-function in Eq. (10) becomes $E_{\mathbf{k}_L} - E_{\mathbf{k}_R} = E_{DL} - E_{DR} + \hbar v_F k_L - \hbar v_F k_R = eV - 2\Delta E + \hbar v_F (k_L - k_R)$. In evaluating Eq. (10) this energy δ-function can be used to eliminate the sum over the $k_R$ magnitude, with $k_R = k_L + eV'/\hbar v_F$ where we have introduced $eV' \equiv eV - 2\Delta E$ (for $V' < 0$, the constraint that $k_R \geq 0$ must explicitly be applied). The current from region III is identical to that from region I. In region II we find $k_R = e|V'|/\hbar v_F - k_L$ with $0 \leq k_L \leq e|V'|/\hbar v_F$.

Considering Eq. (10) in the limit of large *A*, we have $\mathbf{k}_L = \mathbf{k}_R \equiv \mathbf{k}$ since $\Lambda(\Delta \mathbf{k}) \to \delta_{\mathbf{k}_L, \mathbf{k}_R}$, so that the equation becomes

$$I = \frac{8\pi e}{\hbar} \left( \frac{\hbar^2 \kappa e^{-\kappa d}}{2mD} \right)^2 \sum_{B, \mathbf{k}} |g_0(\theta_{\mathbf{k}}, \theta_{\mathbf{k}})|^2 \left[ f_L(E_{L,\mathbf{k}}) - f_R(E_{R,\mathbf{k}}) \right] \delta(E_{L,\mathbf{k}} - E_{R,\mathbf{k}}) \tag{11}$$

where we have added indices *L* and *R* to the energies to make it clear which electrode they are associated with. We note that for tunneling between like valleys and unlike bands, $|g(\theta_{\mathbf{k}}, \theta_{\mathbf{k}})| = 2u_{12}^2 \sin(\theta_{\mathbf{k}})$, with the term involving $u_{11}^2$ having been eliminated. This cancellation occurs because of orthogonality between the lateral portions of the VB and CB wavefunctions, but nevertheless nonzero tunnel current is still produced by the $u_{12}$ cross-term.

In the following Section we evaluate Eq. (10) for large-area rotationally aligned electrodes, and in the Section after that we consider finite-area rotationally aligned electrodes. The case of rotational misalignment is considered in the numerical results of Section III.

**C. Tunneling current for large-area graphene sheets**

In this Section we focus our discussion to large electrode areas with no misorientation between the electrodes ($\mathbf{Q}=\mathbf{0}$). We first consider an undoped GIG junction,[13] the band structure for which is pictured in Fig. 3. Given the requirement of **k**-conservation as enforced by Eq. (9) for large $A$, there is only a single ring of **k**-points that satisfy that, located at an energy midway between the Dirac points as shown in Fig. 3. Thus for $V>0$ we need only consider VB states for the left electrode, $E_{L,\mathbf{k}} = E_{DL} - \hbar v_F k$, and CB states for the right electrode, $E_{R,\mathbf{k}} = E_{DR} + \hbar v_F k$ (or vice versa for $V<0$). Thus, $E_{L,\mathbf{k}} - E_{R,\mathbf{k}} = E_{DL} - E_{DR} - 2\hbar v_F k = eV - 2\hbar v_F k$. Substituting into the δ-function of Eq. (11), and evaluating the sum over **k** as an integral in the usual way, yields the current for tunneling between like valleys,

$$I = \frac{8\pi e}{\hbar}\left(\frac{\hbar^2 \kappa e^{-\kappa d}}{2mD}\right)^2 \frac{A}{2\pi} 2 u_{12}^4 \int_0^{k_{\max}} k\, dk\, \left[f_L(E_{L,\mathbf{k}}) - f_R(E_{R,\mathbf{k}})\right] \delta(e|V| - 2\hbar v_F k). \quad (12)$$

where $k_{\max} = e|V|/\hbar v_F$. For tunneling between unlike valleys the term $u_{12}^2$ is replaced by $u_{11}^2$. The integral is easily evaluated using the δ-function, yielding for zero temperature

$$\begin{aligned}
I &= \frac{8\pi e}{\hbar}\left(\frac{\hbar^2 \kappa e^{-\kappa d}}{2mD}\right)^2 \frac{A u_{12}^4}{\pi}\frac{1}{2\hbar v_F}\frac{eV}{2\hbar v_F} \\
&= \frac{e^2 A}{2\hbar}\left(\frac{\hbar \kappa u_{12}^2 e^{-\kappa d}}{mD v_F}\right)^2 V
\end{aligned} \quad (13)$$

Now let us turn to a doped GIG junction. We first consider the nonresonant case when the band structures are not aligned, i.e. $eV \neq 2\Delta E$, as in Figs. 1(a) or 1(b). The situation then is similar to the undoped junction, with a single ring of **k**-values satisfying wavevector conservation for each particular voltage. The derivation of the tunnel current is very similar to the undoped case. For example, for the situation pictured in Fig. 1(a) we have for the relevant states that $E_{L,\mathbf{k}} = E_{DL} + \hbar v_F k$ and $E_{R,\mathbf{k}} = E_{DR} - \hbar v_F k$ so that $E_{L,\mathbf{k}} - E_{R,\mathbf{k}} = E_{DL} - E_{DR} + 2\hbar v_F k = eV - 2\Delta E + 2\hbar v_F k$. Thus, in Eq. (11) we have, $\delta(E_{L,\mathbf{k}} - E_{R,\mathbf{k}}) = \delta(eV - 2\Delta E + 2\hbar v_F k)$. Therefore the current at zero temperature is given by

$$I = \frac{e^2 A}{2\hbar}\left(\frac{\hbar\kappa u_{12}^2 e^{-\kappa d}}{mDv_F}\right)^2 \left(\frac{2\Delta E}{e} - V\right) \quad (14)$$

for $0 < eV < 2\Delta E$, and by the negative of that for $V < 0$ (since the sign of $f_L - f_R$ changes). Similarly, for voltages of $eV > 2\Delta E$ we have for the relevant states $E_{L,\mathbf{k}} = E_{DL} - \hbar v_F k$ and $E_{R,\mathbf{k}} = E_{DR} + \hbar v_F k$ so that $E_{L,\mathbf{k}} - E_{R,\mathbf{k}} =$ $E_{L,\mathbf{k}} - E_{R,\mathbf{k}} = E_{DL} - E_{DR} - 2\hbar v_F k = eV - 2\Delta E - 2\hbar v_F k$ and $\delta(E_{L,\mathbf{k}} - E_{R,\mathbf{k}}) = \delta(eV - 2\Delta E - 2\hbar v_F k)$. Therefore the current is

$$I = \frac{e^2 A}{2\hbar}\left(\frac{\hbar\kappa u_{12}^2 e^{-\kappa d}}{mDv_F}\right)^2 \left(V - \frac{2\Delta E}{e}\right). \quad (15)$$

Both Eqs. (14) and (15) apply to tunneling between like valleys; for unlike valleys, the term $u_{12}$ is replaced by $u_{11}$.

Now we turn to the resonant situation, with $eV = 2\Delta E$ in the doped GIG junction. As pictured in Fig. 1(c), there are states existing over all energies that satisfy the requirement of **k**-conservation. We have $E_{L,\mathbf{k}} - E_{R,\mathbf{k}} = E_{DL} - E_{DR} = eV - 2\Delta E$ for each pair of states, leading to $\delta(E_{L,\mathbf{k}} - E_{R,\mathbf{k}}) = \delta(0)$ in Eq. (11) which is not well defined. In the following Section we demonstrate how this current *can* be evaluated, first by performing the sums for finite-area graphene sheets using Eq. (10) together with Eq. (9), and then taking the limit of large area. We find an approximate (but fairly accurate) expression for the current as

$$I = \frac{2(0.4)e^2 A}{\sqrt{2\pi}\hbar}\left(\frac{\hbar\kappa e^{-\kappa d}}{mDv_F}\right)^2 \frac{L\Delta E^2(2u_{11}^4 + u_{12}^4)}{e\hbar v_F}\exp\left\{-\frac{A}{4\pi}\left[\frac{(eV - 2\Delta E)}{\hbar v_F}\right]^2\right\}. \quad (16)$$

This equation applies to tunneling between like valleys; for unlike valleys, the $u_{11}$ and $u_{12}$ terms are interchanged.

The occurrence of $L$ in Eq. (16) is worth examining. As derived in the following Section, the value of $L$ is simply the lateral extent of a graphene sheet (i.e. area of $A = L^2$). However, it is also of interest to consider the effect of structural imperfections in the graphene. Let us say that the graphene can be decomposed into small structurally perfect areas, each with area $a = \ell^2$, and say that there are $M$ such areas in the entire sheet so that $A = M a$. The tunnel current from a single perfect section of the film would be given by Eq. (16), but with $A = a$ and $L = \ell$. The current from the entire sheet would then be given by $M$ times that, yielding a result identical to Eq. (16) but with $L = \ell$. Thus, we can take Eq. (16) to apply to the general case, but with $L$ in that equation interpreted as the lateral extent of perfect areas (i.e. grains) in the graphene. We refer to this lateral extent as a *structural coherence length* in the graphene. For a small, perfect graphene flake, $L$ would be the total lateral extent of the flake, but in a larger defective sheet of graphene, $L$ is the lateral extent of structurally perfect grains in the sheet.

## D. Finite-size effects

We consider the situation for finite-sized areas of graphene, extending over $-L/2 < x < L/2$ and $-L/2 < y < L/2$. The factor $\Lambda(\Delta\mathbf{k})$ introduced in Eq. (9) is easily evaluated to be

$$\Lambda(\Delta\mathbf{k}) = \left| \frac{1}{A} \int_{-L/2}^{L/2} dx \int_{-L/2}^{L/2} dy \, e^{i\Delta\mathbf{k}\bullet\mathbf{r}} \right|^2 = \left| \mathrm{sinc}\left(\frac{L\Delta k_x}{2}\right) \mathrm{sinc}\left(\frac{L\Delta k_y}{2}\right) \right|^2 \tag{17}$$

where $\mathrm{sinc}(x) \equiv \sin(x)/x$. This expression for $\Lambda(\Delta\mathbf{k})$ is of course peaked when $\mathbf{k}_L = \mathbf{k}_R$. Substituting this form into Eq. (10), and converting the sums over $\mathbf{k}_L$ and $\mathbf{k}_R$ to integrals, permits numerical evaluation of the tunneling current (both resonant and nonresonant). It is this method that we use for all of the numerical results presented below.

However, with the goal of obtaining analytical formulas for the tunnel current, use of Eq. (17) for $\Lambda(\Delta\mathbf{k})$ is inconvenient since it does not permit explicit evaluation of the integrals. To achieve this goal, we replace $\Lambda(\Delta\mathbf{k})$ by another function that is also peaked when $\mathbf{k}_L = \mathbf{k}_R$,

$$\tilde{\Lambda}(\Delta\mathbf{k}) \equiv \exp\left(-\frac{1}{\pi}\frac{A|\Delta\mathbf{k}|^2}{4}\right) = \exp\left(-\frac{1}{\pi}\frac{A\Delta k_x^2}{4}\right) \exp\left(-\frac{1}{\pi}\frac{A\Delta k_y^2}{4}\right). \tag{18}$$

The factor of $1/\pi$ in the exponents here is chosen such that the area under $\tilde{\Lambda}(\Delta\mathbf{k})$ when integrated over $\Delta k_x$ or $\Delta k_y$ is identical to that under $\Lambda(\Delta\mathbf{k})$. Using $\tilde{\Lambda}(\Delta\mathbf{k})$ rather than $\Lambda(\Delta\mathbf{k})$ now allows us to explicitly evaluate the sums (integrals) over $\mathbf{k}_L$ and $\mathbf{k}_R$ in Eq. (10). Expressing $|\Delta\mathbf{k}|^2 = k_L^2 + k_R^2 - 2k_L k_R \cos\theta$ where $\theta = \theta_L - \theta_R$ is the angle between $\mathbf{k}_L$ and $\mathbf{k}_R$, the angular part of the integrals is given by

$$\int_0^{2\pi} d\theta_L \int_0^{2\pi} d\theta_R |g(\theta_L, \theta_R)|^2 \tilde{\Lambda}(\Delta\mathbf{k}) =$$
$$\exp\left\{-\frac{1}{\pi}\frac{A}{4}\left(k_L^2 + k_R^2\right)\right\} \int_0^{2\pi} d\theta_L \int_0^{2\pi} d\theta_R |g_0(\theta_L, \theta_R)|^2 \exp\left(\frac{1}{\pi}\frac{A}{2} k_L k_R \cos\theta\right). \tag{19}$$

For tunneling between like valleys, the double integral over $\theta_L$ and $\theta_R$ on the right-hand side equals $8\pi^2 [(u_{11}^4 + u_{12}^4) I_0(A k_L k_R / 2\pi) \pm u_{11}^4 I_2(A k_L k_R / 2\pi)]$ where $I_n$ is a modified Bessel function of the first kind of order $n$ and the upper (lower) sign holds for tunneling between like (unlike) bands. For tunneling between unlike valleys the result is the same but with $u_{11}$ and $u_{12}$ interchanged. Substituting into Eq. (10) we have

$$I = \frac{16\pi e}{\hbar} \left( \frac{\hbar^2 \kappa e^{-\kappa d}}{2mD} \right)^2 \frac{A^2}{(2\pi)^2} \sum_B \int k_L dk_L \int k_R dk_R \left[ f_L(E_{\mathbf{k}_L}) - f_R(E_{\mathbf{k}_R}) \right]$$
$$\times \exp\left( -\frac{A(k_L^2 + k_R^2)}{4\pi} \right) \left[ (u_{11}^4 + u_{12}^4) I_0\left( \frac{A k_L k_R}{2\pi} \right) \pm u_{11}^4 I_2\left( \frac{A k_L k_R}{2\pi} \right) \right] \delta(E_{\mathbf{k}_L} - E_{\mathbf{k}_R}). \quad (20)$$

Let us initially consider the resonant case, so that the region II of the band alignment has zero size. The current from regions I and III are equal so that we need only evaluate only one of them, and we use the CBs. The band structures are aligned, so that $\delta(E_{\mathbf{k}_L} - E_{\mathbf{k}_R}) = \delta(\hbar v_F k_L - \hbar v_F k_R) = \delta(k_L - k_R)/\hbar v_F$ and the current reduces to (including a factor of 2 to account for both regions I and III)

$$I = \frac{32\pi e}{\hbar} \left( \frac{\hbar^2 \kappa e^{-\kappa d}}{2mD} \right)^2 \frac{A^2}{(2\pi)^2 \hbar v_F} \int k^2 dk \left[ f_L(E_{L,\mathbf{k}}) - f_R(E_{R,\mathbf{k}}) \right]$$
$$\times \exp\left( -\frac{Ak^2}{2\pi} \right) \left[ (u_{11}^4 + u_{12}^4) I_0\left( \frac{Ak^2}{2\pi} \right) + u_{11}^4 I_2\left( \frac{Ak^2}{2\pi} \right) \right]. \quad (21)$$

For zero temperature the integrals involving the Bessel functions can be explicitly evaluated. Introducing the integration variable $x = Ak^2/2\pi$, we note that

$$\int_0^{x_{\max}} \sqrt{x} \, dx \, e^{-x} I_0(x) = \frac{2}{3} x_{\max}^{3/2} \, {}_2F_2\left( \left\{ \frac{1}{2}, \frac{3}{2} \right\}, \left\{ 1, \frac{3}{2} \right\}, -2x_{\max} \right) \quad (22)$$

and

$$\int_0^{x_{\max}} \sqrt{x} \, dx \, e^{-x} I_2(x) = \frac{1}{28} x_{\max}^{7/2} \, {}_2F_2\left( \left\{ \frac{5}{2}, \frac{7}{2} \right\}, \left\{ \frac{9}{2}, 5 \right\}, -2x_{\max} \right) \quad (23)$$

with $x_{\max} = A k_{\max}^2/2\pi = A(\Delta E/\hbar v_F)^2/2\pi$ and where ${}_2F_2$ is a generalized hypergeometric function. By numerical inspection, we find that the quantities on the right-hand side of the equals sign for *both* Eqs. (22) and (23) approach, for large $x_{\max}$, $(0.399...) x_{\max}$, which we express simply as $0.4 x_{\max}$. We thus obtain a formula for the peak resonant current ($V = 2\Delta E/e$) at zero temperature of

$$I = \frac{32\pi e}{\hbar} \left( \frac{\hbar^2 \kappa e^{-\kappa d}}{2mD} \right)^2 \frac{A}{(2\pi)^2} \sqrt{\frac{\pi}{2}} \frac{L \Delta E^2}{(\hbar v_F)^3} 0.4 (2u_{11}^4 + u_{12}^4) \quad (24)$$

This expression applies to tunneling between like valleys; for unlike valleys, $u_{11}$ and $u_{12}$ are interchanged. In the following Section we compare this result to the numerical evaluation of the current from Eqs. (10) and (17), and we find that they agree fairly well.

Finally, for the current away from the resonant peak, we return to Eq. (20) and evaluate it in the various energy regions of band alignment shown in Fig. 2. In region I we have $k_R = k_L + eV'/\hbar v_F$ with $eV' \equiv eV - 2\Delta E < 0$. In the integrand of Eq. (20) there is the term $\exp[-A(k_L^2+k_R^2)/4\pi]$, which, with $k_R = k_L + eV'/\hbar v_F$, will be sharply peaked at $k_L^2 = k_R^2 = (eV'/2\hbar v_F)^2$. For these $k_L$ and $k_R$ values the argument of the $I_0$ Bessel function will be $>1$ for $V' > (3\text{ V nm})/L$, which corresponds to 0.03 V for $L=100$ nm or 0.003 V for $L=1000$ nm. For these cases we can replace the Bessel function by its asymptotic limit, $\exp(Ak_L k_R / 2\pi)/\sqrt{Ak_L k_R}$. Combining with the $\exp[-A(k_L^2+k_R^2)/4\pi]$ term, and expressing the exponent as $k_L^2 + k_R^2 - 2k_L k_R = (k_L - k_R)^2 = (eV'/\hbar v_F)^2$, we are left with a term $\exp[-A(eV'/\hbar v_F)^2/4\pi]$ which gives the dependence of the current on $V'$. The same term arises when we consider the energy region III, and similar arguments can be made for the $I_2$ Bessel function (albeit for larger $V'$). In both these regions the tunneling occurs between like bands, so the term $2u_{11}^4 + u_{12}^4$ in Eq. (24) is appropriate. Therefore, to provide an approximate analytic expression for the entire (broadened) resonant peak of the current, we simply take the peak value from Eq. (24) and multiply that by $\exp[-A(eV'/\hbar v_F)^2/4\pi]$. The final expression is then listed above in Eq. (16). As shown in the following Section, this approximate expression for the current actually provides quite good results even for $V'$ values that are nearer to zero than by the bounds just stated. For the off-resonance contribution from region II we maintain our usage of Eqs. (14) and (15), with the term $u_{12}^4 = (u_{11}^4 + u_{12}^4) - u_{11}^4$ being appropriate for the unlike bands. It should however be noted that, close to 0 V, Eq. (14) does not properly describe the linear current-voltage relationship that occurs for finite electrode area, as illustrated in the following Section.

**E. Charging of the Graphene Electrodes**

In the derivations of the previous Sections we treated $\Delta E$ (the separation of the Fermi level and Dirac point) as if it were a fixed quantity. However, for any physical GIG junction $\Delta E$ will actually vary with the voltage $V$ between the electrodes due to charging of the graphene electrodes. To illustrate this effect, we consider initially the situation for nominally undoped electrodes as pictured in Fig. 4. If the electrodes were metallic, then a surface charge would form on each electrode in response to the electric field across the junction. For the case of graphene electrodes, this "surface charge" becomes a 2D charge within each electrode. The GIG junction has associated with it a geometric capacitance per unit area, $C = \varepsilon_R \varepsilon_0 / d$, where $\varepsilon_R$ is the relative dielectric constant and $d$ is the thickness of the insulating layer.[7] For a voltage across the insulator of $V_i$, the charge density in the electrodes is given by

$$\sigma = CV_i = e(n_L - p_L) = e(p_R - n_R) \quad (25)$$

where $n$ and $p$ are the 2D carrier densities in the respective electrodes. Here, $V_i$ is the same as $V'$ defined above; we use this new symbol to signify that it is the voltage across the insulator with the graphene electrode quantum capacitance considered.[7] The applied voltage $V$ between the electrodes is given by $eV = \mu_L - \mu_R$.[14] Thus, referring to Fig. 4, we have

$$eV = eV_i + (\mu_L - E_{DL}) + (E_{DR} - \mu_R) \qquad (26)$$

where for the undoped electrodes $(\mu_L - E_{DL}) = (E_{DR} - \mu_R) \equiv \Delta E$. A general expression for the carrier densities is

$$n - p = \frac{2}{\pi(\hbar v_F)^2} \left\{ \int_{E_D}^{\infty} \frac{(E - E_D)\, dE}{1 + \exp[(E - \mu)/k_B T]} - \int_{-\infty}^{E_D} \frac{(E - E_D)\, dE}{1 + \exp[(\mu - E)/k_B T]} \right\}$$

$$= \frac{2}{\pi(\hbar v_F)^2} \left\{ \int_0^{\infty} \frac{E\, dE}{1 + \exp[(E - \Delta E)/k_B T]} - \int_{-\infty}^0 \frac{E\, dE}{1 + \exp[(\Delta E - E)/k_B T]} \right\} \qquad (27)$$

which depends only on $\Delta E = \mu - E_D$. Thus, substituting the expression for $V_i$ from Eq. (25) into Eq. (26), we are left with a single equation for $\Delta E$ that can easily be solved numerically.

Moving to the case of doped electrodes, Eq. (25) becomes generalized to read

$$\sigma = CV_i = e[(n_L - p_L) - N_D] = e[(p_R - n_R) - N_A] \qquad (28)$$

where 2D substitutional doping concentrations of $N_D$ (n-type) in the left-hand electrode and $N_A$ (p-type) in the right-hand electrode are assumed. We consider equal concentrations in both electrodes, $N_D = N_A = N$, so that $(n_L - p_L) = (p_R - n_R)$ and $(\mu_L - E_{DL}) = (E_{DR} - \mu_R) \equiv \Delta E$. Equation (26) still applies, and substituting Eq. (28) into that we arrive at the single equation

$$eV = \frac{e^2}{C}[(n_L - p_L) - N] + 2\Delta E \qquad (29)$$

where $(n_L - p_L)$ is given by Eq. (27). Given $V$, $C$, and $N$, this equation can be solved numerically for $\Delta E$. For zero temperature this solution is easily expressed, with $(n_L - p_L) = \pm \Delta E^2 / [\pi(\hbar v_F)^2]$ where the upper sign if used for $n_L > p_L$ ($\Delta E > 0$) and the lower sign for $n_L < p_L$ ($\Delta E < 0$). Equation (29) then forms a quadratic equation for $\Delta E$, with the solution

$$\Delta E = \pm \frac{1}{2} \left\{ \frac{-2C\pi(\hbar v_F)^2}{e^2} + \sqrt{\left[\frac{2C\pi(\hbar v_F)^2}{e^2}\right]^2 \pm 4\pi(\hbar v_F)^2 \left(N + \frac{CV}{e}\right)} \right\}. \qquad (30)$$

This solution is valid for all values of $V$, with $\Delta E > 0$ for $V > -eN/C$ (upper sign), $\Delta E = 0$ for $V = -eN/C$, and $\Delta E < 0$ for $V < -eN/C$ (lower sign).

Using the value of $\Delta E$ deduced from the above procedure, the tunneling current in the GIG junction can be computed using the formulas of the previous Sections.[15] As an example of the influence of the electrode charging, we consider the variation in $\Delta E$ as a function of $V$ for two situations: one for a thin insulating layer, taking $\varepsilon_R = 4$ and $d = 0.5$ nm which gives a capacitance of $C = 7.1 \times 10^{-20}$ F/nm$^2$, and another for a relatively thick insulator with ten times smaller capacitance. Figure 5 shows the resulting $\Delta E$ values, assuming a doping concentration of $0.74 \times 10^{12}$ cm$^{-2}$ corresponding to a value of $\Delta E = 0.1\,\text{eV}$ for $C = 0$. As can be seen from the plot, the variation in $\Delta E$ for the thick insulator is not particularly large, and as will be seen in the following Section it produces only a modest broadening of the resonant peak in the current. For the thin insulator the variation of $\Delta E$ is much greater, leading to a substantial broadening of the resonant peak in the tunnel current.

## III. Results

In this Section to consider numerical results for the single-particle tunnel current in doped GIG junctions, assuming initially a fixed value of $\Delta E$ for the electrodes (i.e. zero capacitance of the junction). Figure 6 shows results for $\Delta E = 0.1\,\text{eV}$, as given by Eq. (10) for the exact (numerical) solution, at temperatures of $T = 0\,\text{K}$ and 300 K. Also shown are the predictions of our approximate (analytic) formulas for the current, at 0 K, as given by the sum of Eq. (16) with Eq. (14) or (15). These formulas provide a reasonably good description of the current, although they do not capture the asymmetry of the resonance peak (this asymmetry arises from regions I and III of the band alignment, Fig. 2, the current from which has different magnitude for $V > 2\Delta E/e$ or $V < 2\Delta E/e$). There is little temperature dependence in the *width* of the resonant peak, as already noted in prior work,[2,3] although the *height* of the peak increases somewhat with temperature since greater numbers of states are accessed at the higher *T* (temperature dependence of the *I-V* curve is also apparent close to 0 V, with the slope of the *I-V* curve there being affected both by *T* and *L*). As discussed in Section II(B), the height of the resonant peak is proportional to the structural coherence length *L*, with the width being proportional to $1/L$. The nonlinearity of the *I-V* curve is large in Fig. 6, and for larger coherence lengths (and/or larger $\Delta E$) it becomes larger still.

The results in Fig. 6 are applicable to graphene electrodes that have perfectly aligned crystal orientations. For the case of rotational misalignment between the electrodes, we still evaluate the current using Eq. (10), but we now include the $\exp(i\mathbf{Q}\bullet\mathbf{r})$ term in the definition of $\widetilde{\Lambda}(\Delta\mathbf{k})$ [i.e. as in the integral of Eq. (7)]. Results of that type of computation are shown in Fig. 7. As the misalignment angle increases, the intensity of the resonant peak at $V = 2\Delta E/e$ rapidly decreases; the peak shifts to higher voltages and a related peak appears at lower (negative voltages). For the situation of $L = 100\,\text{nm}$ being considered, it is apparent from Fig. 7 that only the graphene grains in the opposing electrodes that are misoriented by less than about ±0.15° will contribute significantly to the resonant peak. Compared to a total angular range of $-30°$ to $+30°$ (beyond which a resonance between the next-nearest valleys, i.e. K and K', must be considered), it is apparent that only 0.5% of the area if each electrode contributes to the resonant peak (i.e. for randomly oriented grains in the electrodes). The other, surrounding graphene grains do nevertheless play an important role of laterally transporting the current. For the larger grain size of $L = 1000\,\text{nm}$, only areas of the opposing electrodes that are misoriented by less than about ±0.015° contribute significantly to the resonant peak, corresponding to 0.05% of the electrode areas.

For the *I-V* characteristics of misaligned electrodes ($\omega > 0.15°$) displayed in Fig. 7, it is apparent that they also have peak currents, but ones that are smaller and at a different voltages than for the aligned case ($\omega = 0°$). These peaks for the misaligned situation arise due to a locus of points in **k**-space where both the wavevectors and the energies of states in the two electrodes are matched, as illustrated in Fig. 8 for $V' > 0$ where $V' \equiv V - 2\Delta E/e$. By inspection, it can be seen that the voltages at which these peaks occur are given by $V' = \pm \hbar v_F Q / e$. The peak currents for the misaligned case become smaller, relative to the peak aligned current, as the structural coherence length *L* increases. However, the range of $\omega$ that contributes to the peak current for aligned electrodes also falls with *L*. The net result is that the peak-to-valley ratio of the angle-averaged current increases sublinearly with *L*, being 1.9 for the *L*=100 nm case of Fig. 7, and 3.7 for *L*=1000 nm. Of course, as *L* increases the total electrode area required such that well-aligned portions of the opposing electrodes will occur also increases, being $\approx L^3/(1 \text{ nm})$ with the assumption of randomly oriented grains in one or both electrodes.

Considering now the effect of the nonzero capacitance of the GIG junction, Fig. 9 displays the resonant peak at zero temperature for the values of capacitance already defined in regard to Fig. 5. The $C = 0$ case pictured there is the same as for Fig. 6 (exact computation). The situation with a relatively thick barrier, having $C = 7.1 \times 10^{-21}$ F/nm$^2$, differs only slightly from the $C = 0$ case. However, for the thin barrier with $C = 7.1 \times 10^{-20}$ F/nm$^2$, the resonant peak is now substantially broadened and also shifted to higher voltages. Nevertheless, a large nonlinearity in the *I-V* characteristic remains, and qualitatively the behavior is the same as for the cases with lower capacitance. Approximate solutions for the tunnel current as given by Eqs. (14) – (16) together with Eq. (30) are not shown in Fig. 8, but they do follow the exact curves quite closely for all values of *C*.

## IV. Discussion

The nonlinear *I-V* curves predicted in this work for GIG junctions occur only when the graphene electrodes have differing chemical potentials, arising from different doping concentrations (i.e. in the same manner as for prior work on 2D-2D tunneling).[2,3] Doping of graphene can be accomplished by a variety of means,[16,17,18,19] and chemical potentials shifted by 0.1 eV or more from the Dirac point, both n-type and p-type, are not uncommon. In this respect the simulations presented here appear to be applicable to physically realizable situations.

It is apparent by comparing Figs. 6 and 7 that a much greater nonlinearity of the *I-V* curve for a doped GIG junction occurs when the electrodes are perfectly rotationally aligned (or with misalignment angle of 60°). This rotational alignment imposes a significant constraint on the devices (one that is not present for the BiSFET devices, as discussed in Section I).[5] The manner in which a rotationally aligned GIG junction will be achieved is not clear at present, since it seems to be incompatible with the exfoliation and transfer type of techniques commonly used in handling graphene flakes.[20] A method more consistent with the requirement of rotational alignment would be direct epitaxy of the graphene electrodes and the insulator. Recent works with BN (an insulator with band gap of 6.0 eV),[21] which can be grown epitaxially,[22] provide key steps in this direction but much work on the epitaxy of 2D materials remains to be done.

Even in the absence of perfect rotational-alignment of the electrodes, a moderate degree of nonlinearity of the *I-V* curve (peak-to-valley ratio ≳2) can still be achieved so long as one or both electrodes consist of small, randomly oriented graphene domains with domain size ≳100 nm. The resonant portion of the current will flow through the small portions of the opposing electrodes that are rotationally aligned, with the remainder of the electrodes serving to connect these "hot spots" and also contributing their own background (non-resonant) current. Graphene grown epitaxially on metal substrates consists typically of micrometer-size constant-thickness domains,[23,24,25] with grain size >50 nm and considerable rotational disorder of the grains,[23] although further quantitative evaluation of that is needed. Graphene grown in vacuum on the C-face of SiC{0001} has ≈50 nm size domains also with considerable rotational disorder, although this disorder only extends over about 10% of the total possible range of rotational angles (judging from the width of the diffraction streaks that extend over ≈3° of a 30° sector).[26,27]

To fully exploit the nonlinear *I-V* curve found for the doped GIG tunnel junction, it is desirable to fashion it into some sort of three-terminal device. This can be accomplished simply by putting the GIG junction between two additional gate electrodes, in a geometry identical to that used in the BiSFET[5] (or, with chemical doping of the GIG electrodes as described above, then just a single gate electrode above or below the GIG junction would likely suffice). With the voltage bias in the GIG junction set to the resonance, then a voltage difference across the gate electrode(s)s will swing the current off resonance and thus achieve amplification of the signal to the gate.

Further comparing the BiSFET operation with the single-particle tunneling effects considered here, we note that the BiSFET, in addition to having a critical temperature below which it must be operated, also relies upon a critical current for its nonlinear response.[5] This critical current would presumably require rather tight tolerances on the insulating layer separating the electrodes in order to achieve good device-to-device reproducibility in the operating voltage. The single-particle tunneling does not have that sort of requirement; the tunnel currents will of course scale with the thickness and barrier height of the tunneling layer, but the operating voltage is only weakly dependent on that, being determined primarily by the relative doping of the two GIG electrodes for low capacitance of the junction and varying slightly (Fig. 9) for high values of the capacitance.

**V. Summary**

In summary, we have computed the single-particle tunneling characteristic for a graphene-insulator-graphene junction with complementary doping in the graphene electrodes. A highly nonlinear *I-V* characteristic is found, with a resonant peak whose width is independent of temperature. The dependence of the tunneling current on both the lateral graphene size of the graphene and the relative rotational orientation of the electrodes is considered. The greatest amount of nonlinearity in the *I-V* characteristic is achieved with nearly perfect rotational orientation of electrodes, which presents a significant challenge in fabrication of such devices. A three-terminal device can be fashioned using additional gate electrode(s) above and/or below the GIG junction, in the same geometry as for the recently proposed BiSFET device.[5]

**Acknowledgements**

This work was supported by the National Science Foundation, grants DMR-0856240 and ECCS-0802125, and the SRC NRI MIND project.

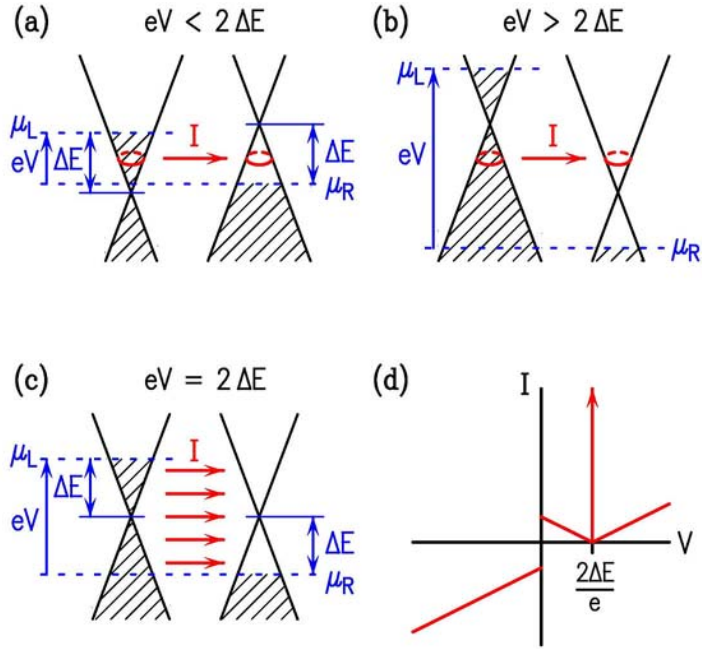

FIG 1. (a) – (c) Band diagrams for a doped GIG junction, at voltages of (a) $V < 2\Delta E/e$, (b) $V > 2\Delta E/e$, and (c) $V = 2\Delta E/e$. In (a) and (b), states satisfying **k-**conservation (i.e. in limit of large electrode area) are shown by the rings located at an energy midway between the Dirac points for the two electrodes. In (c), states at all energies satisfy **k-**conservation. (d) Qualitative current-voltage (*I-V*) characteristic.

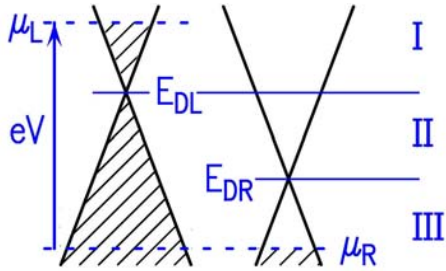

FIG 2. Various energy ranges I, II, and III in a doped GIG junction that must be considered when computing the tunnel current.

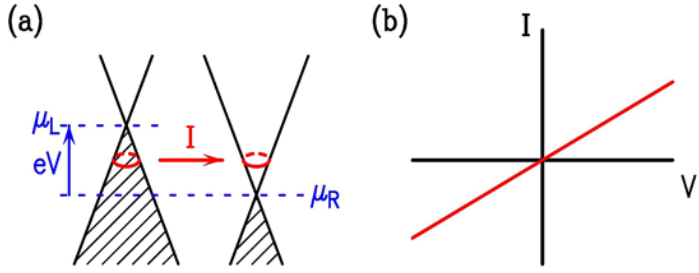

FIG 3. (a) Band diagram for an undoped GIG junction, with states satisfying **k**-conservation (i.e. in limit of large electrode area) shown by the rings located at an energy midway between the Dirac points for the two electrodes. (b) Qualitative *I-V* curve.

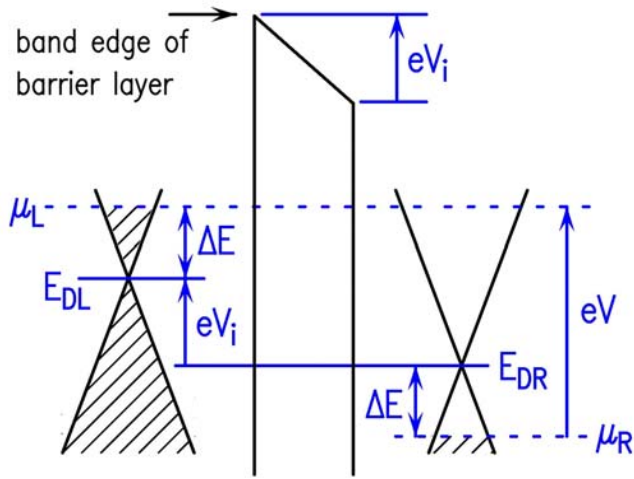

FIG 4. Band diagram for GIG junction with undoped electrodes, including consideration of the capacitance of the insulator layer. Charging of the electrodes results, so that the voltage drop across the insulator $V_i$ is different than the applied voltage $V$ between the electrodes.

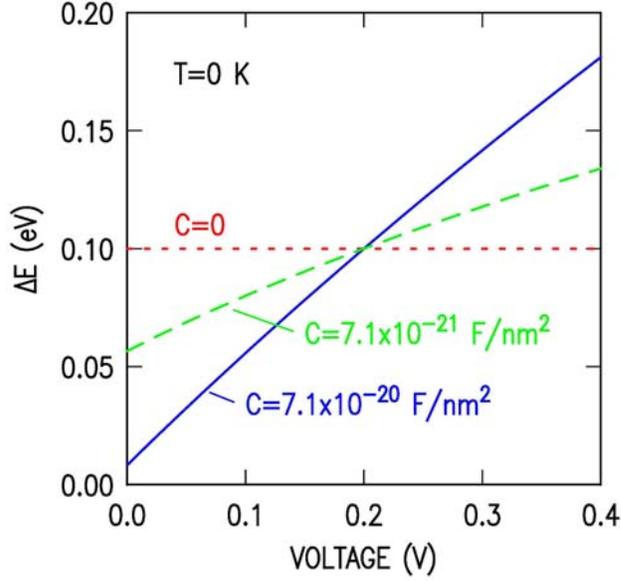

FIG 5. Dependence of $\Delta E$ (the separation of the Fermi level and the Dirac point) on the capacitance of the junction $C$ and the applied voltage $V$ between the graphene electrodes, for a doping concentration corresponding to $\Delta E = 0.1\,\text{eV}$ at zero capacitance.

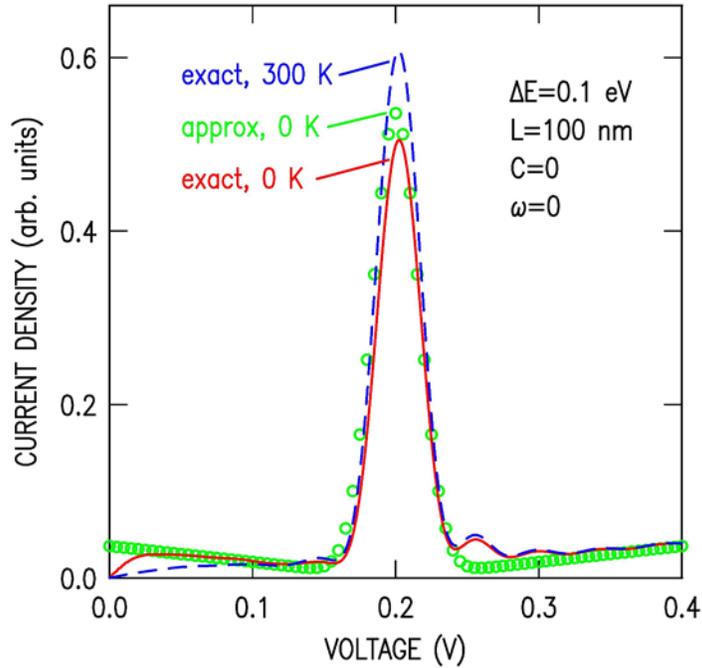

FIG 6. Current vs. voltage in a doped GIG junction, for an energy difference $\Delta E$ between the Fermi-level and the Dirac point in each electrode of 0.1 eV (zero capacitance of junction), and for a structural coherence length of $L = 100\,\text{nm}$. Values of $u_{11} = 1$ and $u_{12} = 1$ are assumed, the graphene lattices in the two electrodes are rotationally aligned.

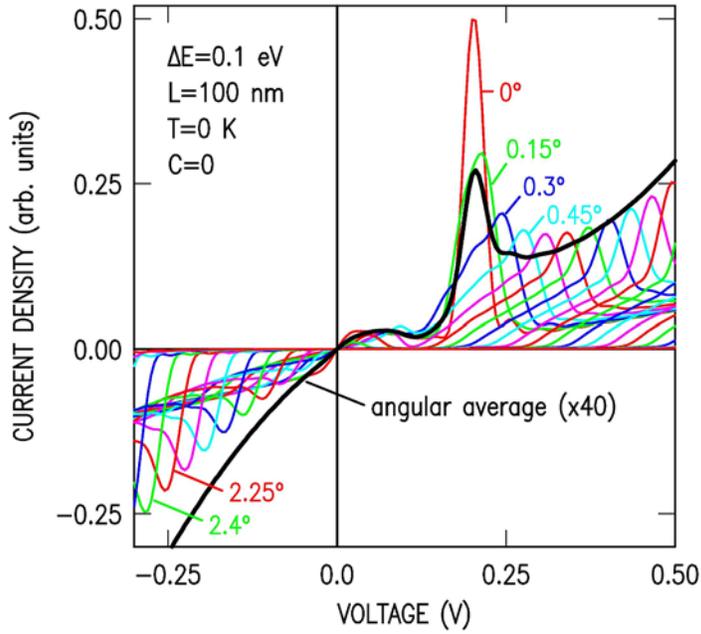

FIG 7. Current vs. voltage in a doped GIG junction with rotationally misaligned electrodes. Individual curves with misalignment angles $\omega$ spaced by 0.15° are shown, with the angular average shown by the thick curve. Results are for an exact computation at 0 K, with other parameters being the same as in Fig. 6.

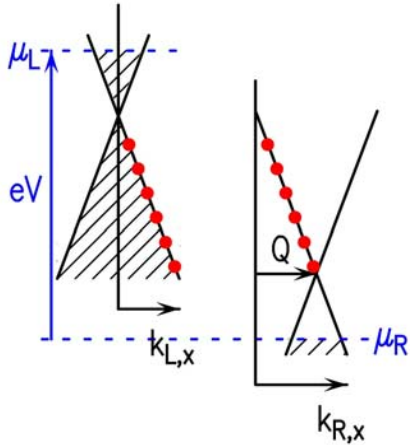

FIG 8. Schematic energy vs. wavevector band structures, illustrating the source of the main component of the tunnel current for rotationally misaligned electrodes. The bands of the right-hand electrode are shifted by a wavevector Q (assumed to be in the x-direction) compared to those of the left-hand electrode. The points indicated by solid dots on the respective band structures have matching wavevector and energy, hence making a relatively large contribution to the current.

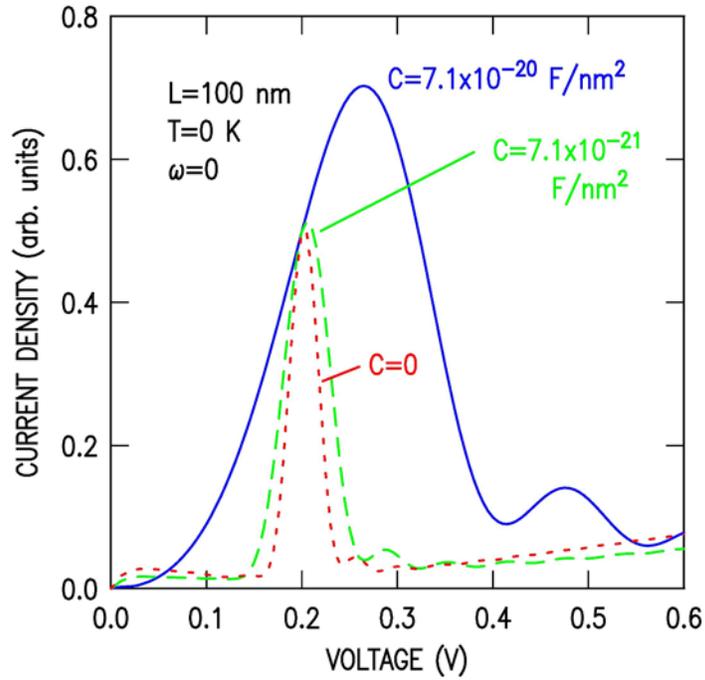

FIG 9. Current vs. voltage in a doped GIG junction, for an exact computation at zero temperature with rotationally aligned electrodes and using a doping concentration that corresponds to $\Delta E = 0.1\,\text{eV}$ at zero capacitance. Various values of the capacitance are considered, with the $\Delta E$ values at each voltage computed as shown in Fig. 5.